\documentclass[runningheads]{llncs}

\usepackage[T1]{fontenc}
\usepackage{pifont}
\usepackage{amssymb}
\usepackage{amsmath}
\usepackage{bbding}

\usepackage{graphicx} %

\usepackage{booktabs}
\usepackage{tabularx}
\usepackage{makecell}
\newcolumntype{P}[1]{>{\centering\arraybackslash}p{#1}}
\newcolumntype{Y}{>{\centering\arraybackslash}X}
\usepackage{threeparttable}
\usepackage[disable]{todonotes}
\usepackage{xcolor}
\usepackage{float}
\usepackage{enumitem}
\usepackage{pdflscape}
\usepackage{afterpage}
\usepackage{csquotes}
\usepackage{enumitem}
\usepackage[hyphens]{url}
\usepackage[hidelinks, breaklinks=true]{hyperref}
\usepackage[style=lncs, urldate=iso]{biblatex}
\DeclareDelimFormat[textcite]{namelabeldelim}{\addnbspace}
\usepackage{tikz}
\usetikzlibrary{math}
\usetikzlibrary{shapes}
\usepackage{pgfplots}

\usepackage[capitalize,nameinlink]{cleveref}
\usepackage{colortbl}
\usepackage[acronym,section]{glossaries}
\glsdisablehyper %

\usepackage{subcaption}
\usepackage{calc}
\usepackage{amstext}
\usepackage{algorithm2e}
\usepackage[bottom]{footmisc}

\makeatletter
\newcommand{\newabbr}[2]{%
  \lowercase{\newabbr@define{#1}}{#1}{#2}%
}
\newcommand{\newabbr@define}[3]{%
  \newacronym{#1}{#2}{#3}%
  \expandafter\newcommand\csname #1\endcsname{\gls{#1}}%
}
\makeatother
\newabbr{PQC}{Post-Quantum Cryptography}
\newabbr{CRQC}{Cryptanalytically Relevant Quantum-Computer}
\newabbr{CARAF}{Crypto Agility Risk Assessment Framework}
\newabbr{KPI}{Key Performance Indicator}
\newabbr{CMMC}{Cybersecurity Maturity Model Certification}
\newabbr{NIST}{National Institute of Standards and Technology}
\newabbr{CSF}{Cyber Security Framework}

\newacronym{dp}{DP}{Design Principle}
\newcommand{\DP}{\gls{dp}}
\newcommand{\DPs}{\glspl{dp}}

\newcommand{\camm}{CAMM}

\newcommand{\cmark}{\ding{51}}
\newcommand{\xmark}{\ding{55}}
\newcommand{\mmark}{\raisebox{0.35ex}{$\thicksim$}}

\newcommand{\abbratis}{ATIS}
\newcommand{\modelatis}{Strategic Framework for Crypto Agility and Quantum Risk Assessment (\abbratis)}
\newcommand{\abbrfsisac}{FS-ISAC}

\tikzset{cross/.style={cross out, draw=black, minimum size=2*(#1-\pgflinewidth), inner sep=0pt, outer sep=0pt},
cross/.default={1pt}}

\renewcommand{\thead}[1]{\textbf{#1}}

\newcommand{\modelfsisac}{Building Cryptographic Agility in the Financial Sector (\abbrfsisac{})}%
\newcommand{\abbrdeloitte}{Deloitte}
\newcommand{\modeldeloitte}{Cryptographic Resilience Community Profile (\abbrdeloitte)}

\begin{document}

\title{Maturity$^2$: Assessing the Maturity of Crypto Agility Maturity Models}
\titlerunning{Maturity$^2$: Assessing the Maturity of Crypto Agility Maturity Models}

\author{Leonie Wolf\inst{1,3,4}\orcidID{0009-0008-1554-9975} \and
Robin Leander Schröder\inst{1,2,3}\orcidID{0000-0002-0791-5700} \and
Sebastian Schinzel\inst{1,3,4,5}\orcidID{0000-0002-7944-5488} \and
Fabian Ising\inst{1,3,4}\orcidID{0000-0001-9852-1231}}

\institute{Fraunhofer SIT
\and Fraunhofer Austria
\and National Research Center for Applied Cybersecurity ATHENE 
\and HNFIZ Cybersecurity 
\and FH Münster University of Applied Sciences  \\ 
\email{\{leonie.wolf,leander.schroeder,sebastian.schinzel,fabian.ising\}\\@sit.fraunhofer.de}
}

\date{January 2026}

\maketitle 

\begin{abstract}
Crypto agility maturity models aim to support organizations in preparing for cryptographic transitions, such as migrating to post-quantum cryptography. Recognizing the importance of measurable maturity, NIST references several such models in its guidance. Given NIST’s influence on cybersecurity practice, these models may shape organizational planning and governance mechanisms. However, their design quality has not been systematically evaluated.\\%
In this work, we assess three industry-driven maturity models against established design principles. Our analysis reveals recurring weaknesses, particularly the lack of measurable, reproducible assessment criteria and limited prescriptive guidance for improving crypto agility. At the same time, we highlight individual strengths that can assist practitioners in choosing among the models and derive concrete recommendations for strengthening future crypto agility maturity models.%
\end{abstract}

\glsresetall

\section{Introduction}

Potential \glspl{crqc} threaten the security of classical asymmetric cryptographic primitives. This threat has spurred the field of \pqc{}, resulting in the largest cryptographic migration in history. The scale, organizational complexity, and partial lack of PQC-ready protocol implementations pose significant challenges for affected organizations. Cryptographic agility (\enquote{crypto agility}) has emerged as a key capability for addressing these challenges, enabling organizations to transition to \pqc{} and systematically replace cryptographic mechanisms in response to future vulnerabilities. Consequently, the ability to manage cryptographic transitions must be deliberately planned and assessed.

Crypto agility maturity models aim to provide standardized frameworks for evaluating and improving their ability to transition among cryptographic algorithms. Recognizing this need, the \nist{} recently published a crypto agility white paper~\cite{nistConsiderations2025} that references four maturity models. The most prominent academic contribution among them is the Crypto Agility Maturity Model (CAMM) by \textcite{hohmCamm2023}, while the remaining three models originate from industry initiatives~\cite{fsisacAgility2024,atisAgility2024,deloitteAgility2025}.
While the white paper cites these models only as illustrative industry initiatives, \nist{}’s role in cybersecurity guidance and procurement practices means such references may come to shape guidance and governance, similar to the requirement of the \cmmc{} in US DoD procurements~\cite{dod-cmmc}. Notably, NIST already recommends including crypto agility as an evaluation criterion for project proposals~\cite{nistConsiderations2025}.

If maturity models are to guide organizational planning or inform governance instruments, their design quality becomes critical. However, aside from the \camm{}, the referenced models have not been independently assessed against established maturity-model design principles. In 2025, \textcite{wolfEval2025} evaluated the CAMM and identified significant shortcomings, including insufficient intersubjective measurability of assessment criteria and a lack of prescriptive guidance for improvement.
Such limitations undermine consistent evaluation and hinder structured advancement of crypto agility.

In this paper, we conduct the first comparative evaluation of the three NIST-cited industry-driven models and compare their design profiles to the existing CAMM evaluation. We seek to answer the following research questions:

\begin{itemize}[align=left, widest={\textbf{RQ1:}},leftmargin=*]
    \item[\textbf{RQ1:}] To what extent do the crypto agility maturity models referenced by NIST fulfill established design principles for maturity models?
    \item[\textbf{RQ2:}] Which weaknesses are systematic shortcomings across the models, and which can be attributed to differing intended purposes?
    \item[\textbf{RQ3:}] How can practitioners select among the models, and what should developers of future models consider?
\end{itemize}

\noindent\textbf{Contributions.}
In this work, we make the following contributions:
\begin{itemize}
    \item We provide the first comparative assessment of the three industry-driven frameworks referenced in the NIST crypto agility white paper~\cite{nistConsiderations2025} and synthesize the results with the existing CAMM evaluation~\cite{wolfEval2025}.
    \item We show which shortcomings are systematic across the models---most critically the absence of intersubjectively verifiable criteria---and which merely reflect their differing purposes.
    \item We offer practitioners guidance for selecting among the models.
    \item We derive recommendations for future
    crypto agility maturity models.
\end{itemize}

\section{Background and Related Work}

\subsection{Maturity Model Design Principles}
Maturity models are widely used to assess organizational properties or capabilities. 
Maturity models can serve either a descriptive purpose by supporting measurement, comparison, and communication of maturity or a prescriptive purpose by guiding improvements in maturity~\cite{poppelbuss2011}.
In their seminal work, Pöppelbuß and Röglinger~\cite{poppelbuss2011} propose a set of \DPs{} to support the development and evaluation of maturity models.
These Design Principles are divided into three categories: (i) Basic Design Principles, (ii) Design Principles for maturity models for descriptive use, and (iii) Design Principles for prescriptive use. 

The Basic Design Principles assess a model's answers to questions such as: Who can use the maturity model? What kinds of maturity can be assessed with it? For what class of entities can it be used? For what purpose is it meant? And how is it designed and documented?

The Design Principles for descriptive maturity models focus on the assessment itself: Is the assessment process well-defined? What advice and guidance are provided for the assessment? And are the results intersubjectively verifiable?

The Design Principles for prescriptive maturity models add improvement criteria. Does the maturity model provide measures of improvement and guidance for decision-making?
According to \textcite{poppelbuss2011}, a maturity model for descriptive use should also fulfill the Basic Design principles, and a prescriptive maturity model should fulfill all Design Principles.

Importantly, \textcite{poppelbuss2011} clearly stated that not every maturity model must meet all the Design Principles. Rather, they can serve as a tool for comparing maturity models and identifying their strengths and weaknesses.

\subsection{Crypto Agility}
In its most basic form, cryptographic agility (crypto agility) refers to the ability to switch among different cryptographic primitives with minimal effort.
\textcite{nather2024toward} provide a more comprehensive definition:
\enquote{%
Cryptographic Agility is a theoretical or practical approach, objective, or property that provides capabilities for setting up, identifying, and modifying
encryption methods and keying material in a flexible and efficient way while preserving business continuity.}
Unless otherwise noted, we use this definition when discussing crypto agility.

The increasing urgency of crypto agility is closely tied to the potential emergence of \glspl{crqc} and the migration to \pqc{}.
In preparation for this migration, public authorities began advocating for crypto agility \cite{chen_report_2016,eu-pqc-roadmap,BSI-TR-02102-1}.
In 2025, NIST
published a white paper on crypto agility \cite{nistConsiderations2025}, surveying current approaches, outlining challenges and possible trade-offs, and identifying working areas that require additional consideration, i.e., maturity models.

To this end, \textcite{hohmCamm2023} introduced the \enquote{Crypto-Agility Maturity Model (\camm{})}, comprising an extensive list of requirements ranging from technical hardware and software requirements to organizational process requirements. 
More recently, several industry-driven frameworks~\cite{atisAgility2024,fsisacAgility2024,deloitteAgility2025} have been published. As the primary targets of the evaluation in this work, they are introduced in greater detail in the next section.

\subsection{Related Work}

Most notably, \textcite{wolfEval2025} evaluated the CAMM~\cite{hohmCamm2023} using the design criteria proposed by \textcite{poppelbuss2011} and applied it to a common web application scenario.
We extend this line of work by assessing the three remaining crypto agility maturity models~\cite{fsisacAgility2024,atisAgility2024,deloitteAgility2025} referenced by NIST~\cite{nistConsiderations2025}.

Several publications address PQC migration and crypto agility more broadly.
\textcite{nethenPQC2024} design a PQC migration process that they claim fulfills the requirements of the CAMM~\cite{hohmCamm2023}, whereas \textcite{natherMigrating2024} systematically review literature concerned with \pqc{} migrations. Several governmental organizations have recently published guidance on migration to \pqc{}~\cite{eu-pqc-roadmap,bsiSecuring2025,tnoPQCMigration2025}. In 2025, \textcite{jena_quantum_security_deadline_2025} presented a roadmap for implementing crypto agility and recommended policy and compliance measures for PQC readiness. A recent preprint~\cite{erol_quantum_readiness_cryptography_2025} proposes a \enquote{Quantum Readiness Maturity Model} that assesses technical and organizational dimensions in five levels of maturity. 
The \caraf{}~\cite{caraf2021}, focuses specifically on assessing and mitigating risks associated with insufficient agility.

Other studies have proposed new, domain-specific criteria for evaluating Business Process Maturity Models~\cite{felch2020make,looyEvaluating2017} and Industry 4.0 Maturity Models~\cite{dikhanbayeva2020assessment}. 
\textcite{roglinger2012maturity} evaluated Business Process Management tools by applying and further developing the Design Principles laid out by \textcite{poppelbuss2011}. \textcite{frick_identification_design_elements_2012} used the maturity model design principles as one of two frameworks to analyze 23 maturity models in the domain of interorganizational systems. \textcite{ferraz_bim_maturity_models_2020} built on the design principle to evaluate Building Information Modeling maturity models. These works demonstrate the broader applicability of structured design principles for assessing maturity models.

\section{Assessment Methodology}
We comparatively assess the three previously unevaluated industry-driven frameworks referenced by NIST using established design principles~\cite{poppelbuss2011} for maturity models. We then contrast the results with the earlier CAMM evaluation by \textcite{wolfEval2025} to distinguish recurring shortcomings from properties attributable to model purpose and structure.

\subsection{Models under Test}%

The NIST white paper on crypto agility~\cite{nistConsiderations2025} references four crypto agility maturity models: the \camm{}~\cite{hohmCamm2023}, a model by the Financial Services Information Sharing and Analysis Center ({FS-ISAC})~\cite{fsisacAgility2024}, one by the Alliance for Telecommunications Industry Solutions (ATIS)~\cite{atisAgility2024}, and one by Deloitte~\cite{deloitteAgility2025}. 
Since the \camm{}~\cite{hohmCamm2023} has already been evaluated by \textcite{wolfEval2025}, we focus on the remaining three models. In the following, we refer to the models as \abbrfsisac{}, \abbratis{}, and \abbrdeloitte{}.

The section \enquote{Maturity Assessment for Crypto Agility} in the NIST white paper could also be interpreted as a preliminary maturity model based on the NIST Cybersecurity Framework (CSF) 2.0 \cite{nistcsf2024}. However,  NIST explicitly frames it as elements that \enquote{a crypto agility maturity model might include} and it spans fewer than two pages. Given its high-level nature, we do not consider it a fully specified maturity model ready for full-scale evaluation.

We evaluate the models referenced by \nist{}, as we expect these to gain the most visibility among practitioners, warranting systematic analysis. To our knowledge, no further crypto agility maturity model has been proposed. Related work either targets the broader domain of quantum readiness (e.g., the Quantum Readiness Assurance Maturity Model~\cite{qramm2025} and the PQC Maturity Model~\cite{pqcmm2025}), targets agility at the application level while explicitly rejecting a single linear maturity hierarchy in favor of orthogonal, non-hierarchical dimensions~\cite{rameshan2026assessment}, or, like the \caraf{}~\cite{caraf2021}, targets risk assessment rather than maturity.

The three models we analyze differ substantially in their design approaches:
\abbrfsisac{} extends the \camm{}, \abbratis{} adopts a quantitatively oriented \kpi{}-driven framework, and \abbrdeloitte{} builds upon the established \nist{}~\csf{}~2.0 Core.
Key aspects of the models are outlined in \cref{tab:comparison}.

\begin{table*}[tb]
\setlength{\tabcolsep}{4pt}
    \centering
    \small
    \caption{Key characteristics of the analyzed models.}
    \label{tab:comparison}
    \rowcolors{1}{white}{gray!10}
    \begin{tabular}{l l l l}
        \toprule
         \thead{Model} & \thead{Basis} & \thead{Published} & \thead{Core elements} \\
         \midrule
         \abbrfsisac{} & \camm{}~\cite{hohmCamm2023} & Oct 2024 & Framework and Levels \\
         \abbratis{} & $-$ & Jan 2024 & \glspl{kpi} \\
         \abbrdeloitte{} & NIST~CSF~2.0~\cite{nistcsf2024} & Apr 2025 & CSF Community Profile \\
         \bottomrule
    \end{tabular}
\end{table*}

\subsubsection{\modelfsisac.}
The document defining this model was published in 2024 by the Post-Quantum Cryptography Working Group of the Financial Services Information Sharing and Analysis Center~\cite{fsisacAgility2024}. 
The first part of the document motivates crypto agility by introducing the threat of \crqc{} and tracing the historical evolution of cryptography.  It further presents a \enquote{framework for replacing an insecure algorithm}, discusses \enquote{challenges to crypto agility migration}, and outlines nine high-level \enquote{Core Elements of a Successful Crypto Agility Transition}.

The second, more technical, part is an adaptation of the \camm{}~\cite{hohmCamm2023}. The authors adopt CAMM's five maturity levels, ranging from \enquote{Level 0 -- Initial/Not possible} to \enquote{Level 4 - Sophisticated/Adaptive}, including the associated requirements with a short description\footnote{Algorithm Exclusion (R2.3) is mentioned only in the level 2 heading but not in the requirements. Since it is not discussed elsewhere, this may be an editorial oversight.}. %
For each level of maturity, the document describes \enquote{Key Characteristics} and \enquote{Roles and Responsibilities}, which appear to constitute the primary extension to the CAMM. %

In addition, the document is augmented with technical \enquote{Process Guidance} and an appendix containing key definitions of crypto agility-related concepts. %
Overall, the document describes several components, each with a distinct focus on crypto agility. The relationship between these elements is not explicitly defined; it is unclear how the different parts are intended to integrate into a single framework or maturity model.

Although the authors do not claim to present a standalone maturity model, NIST cites it as an industry initiative on crypto agility maturity models~\cite{nistConsiderations2025}. Accordingly, we will evaluate it as such in this paper.

\subsubsection{\modelatis.}
In 2024, the Alliance for Telecommunications Industry Solutions (ATIS) published a white paper on crypto agility~\cite{atisAgility2024}.%

The white paper provides an extensive introduction to cryptographic fundamentals and the threat posed by \crqc{}. It also provides a high-level introduction to crypto agility and outlines its main drivers and challenges. ATIS's main contribution is a set of 23 crypto-agility-related metrics. These are grouped into metrics for organizational and operational strategy, software products/services, and hardware products/services. These metrics are mapped to 11 high-level \glspl{kpi}.
Each metric includes a title and definition, a rationale explaining its importance, a mapping to one or more \glspl{kpi}, and guidance on calculation. It concludes with an example of a scorecard-based assessment and general recommendations for transitioning toward crypto agility.

We note that the term \enquote{KPI} is not used consistently throughout the document. In some cases, it refers to the high-level \gls{kpi} list, while in others (e.g., the scorecards), it appears to describe the individual metrics themselves, which may lead to ambiguity in interpretation.

\glsreset{nist}
\glsreset{csf}
\subsubsection{\modeldeloitte.}
The document by \textcite{deloitteAgility2025} is marked as an initial draft and builds upon the \nist{} \csf{} 2.0~\cite{nistcsf2024}. The CSF 2.0 comprises (i) a \textit{core} that organizes cybersecurity outcomes into the functions govern, identify, protect, detect, respond, and recover, further subdivided into categories and subcategories; (ii) four \textit{tiers} that describe increasing levels of risk management maturity; and (iii) \textit{profiles} to model either current or target cybersecurity posture based on selected core elements.

Deloitte’s primary contribution is a CSF Community Profile focused on \enquote{cryptographic resilience}. Community profiles represent shared priorities across organizations and can serve as a foundation for developing organization-specific profiles. In particular, the authors identify  CSF subcategories relevant to cryptographic resilience and supplement with \enquote{Recommendations and Considerations for Cryptographic Resilience}. Furthermore, they provide a brief introduction to cryptographic resilience and the potential risks posed by quantum computers. A short definition of crypto agility is given in the footnotes.

Deloitte treats crypto agility as one component within the broader context of cryptographic resilience.
As a community profile, the document references the CSF multiple times. In our evaluation, we focus on the community profile as the maturity model component, rather than on the CSF 2.0 as a whole.

\subsection{Method for Evaluation}
For this evaluation, we use the Design Principles for maturity models proposed by \textcite{poppelbuss2011}.
These Design Principles are established in the literature for evaluating maturity models~\cite{roglinger2012maturity, wolfEval2025}. 
They comprehensively cover the information a model provides, its representation of maturity, its assessment procedure, and its guidance for improvement, thereby matching both the descriptive and prescriptive purposes claimed by the models. 
In addition, this method enables comparison with the CAMM evaluation conducted by \textcite{wolfEval2025}.
To perform our analysis of the maturity models, for each of the maturity models under evaluation, two researchers
evaluated each of the Design Principles in joint discussion.
To minimize the potential for differing interpretations of the Design Principles, additional literature on the Design Principles~\cite{roglinger2012maturity} and comparisons with other maturity models were considered, as detailed in the evaluation.

\section{Evaluation Results}%
\label{sec:findings}

Applying the \DPs{} of \textcite{poppelbuss2011}, we find that the models diverge on the basic principles (DP~1.x), but none fully satisfies the descriptive (DP~2.x) and all fail the prescriptive (DP~3.x) principles. We interpret the divergence on basic principles as primarily reflecting the models' intended purposes rather than differences in quality. \Cref{tab:results} records the full per-principle verdicts; we refer to our supplemental material~\cite{per_principle_eval} for per-principle explanations of the ratings. Below, we combine them into three overarching themes: one for divergence and two for shared failures. For the observations regarding the \camm{}, we refer to \textcite{wolfEval2025}. Note that they deemed the prescriptive principles not applicable to the \camm{}; we nevertheless record them as not fulfilled in \cref{tab:results} so that all models are compared against the full set of principles.

\afterpage{    \clearpage%
\definecolor{okBlue}{RGB}{0,114,178}
\colorlet{rNo}{black!5}       %
\colorlet{rPart}{okBlue!20}   %
\colorlet{rYes}{okBlue!45}    %
\newcommand{\cNo}{\cellcolor{rNo}\xmark}
\newcommand{\cPart}{\cellcolor{rPart}\mmark}
\newcommand{\cYes}{\cellcolor{rYes}\cmark}
\newlength{\saveaboverule}\newlength{\savebelowrule}
\newcommand{\catrow}[1]{%
    \noalign{\global\saveaboverule=\aboverulesep \global\savebelowrule=\belowrulesep
             \global\belowrulesep=0pt}%
    \midrule
    \rowcolor{gray!15}\multicolumn{6}{l}{\textbf{\textsc{#1}}}\\
    \noalign{\global\aboverulesep=0pt \global\belowrulesep=\savebelowrule}%
    \midrule
    \noalign{\global\aboverulesep=\saveaboverule}%
}
\begin{landscape}
\begin{table}[p]
    \caption{%
Summary of the evaluation of each crypto agility maturity model against the design principles for maturity models.
    Evaluation results for \camm{}~\autocite{hohmCamm2023} are sourced from \textcite{wolfEval2025}; they deem the prescriptive principles not applicable to the \camm{}, which we record as not fulfilled. \xmark{}: Not fulfilled, \mmark{}: Partially fulfilled, \cmark{}: Fully met.
    }
    \label{tab:results}
\small%
\centering%
\renewcommand{\arraystretch}{0.85}%
\renewcommand{\tabularxcolumn}[1]{m{#1}}%
    \begin{tabularx}{\linewidth}{@{} >{\raggedleft\arraybackslash}m{6mm} >{\raggedright\arraybackslash}m{12cm} Y Y Y Y @{\hspace{\tabcolsep}}}%
    \toprule
    \multicolumn{2}{l}{\thead{Design Principle \cite{poppelbuss2011}}} & \thead{\makecell{CAMM\\\cite{hohmCamm2023}}} & \thead{\makecell{\abbrfsisac{}\\\cite{fsisacAgility2024}}} & \thead{\makecell{\abbratis{}\\\cite{atisAgility2024}}} & \thead{\makecell{\abbrdeloitte{}\\\cite{deloitteAgility2025}}} \\
    \catrow{Basic}
    \textbf{1.1} & \textbf{Basic information} & \\
    & a) Application domain and prerequisites for applicability & \cPart & \cPart & \cPart & \cPart \\
    & b) Purpose of use & \cPart & \cYes & \cYes & \cYes\\
    & c) Target group & \cNo & \cYes & \cPart & \cPart \\
    & d) Class of entities under investigation & \cNo & \cPart & \cPart & \cPart \\
    & e) Differentiation from related maturity models & \cYes & \cYes\ & \cNo & \cYes \\
    & f) Design process and extent of empirical validation & \cYes  & \cNo & \cNo & \cNo \\
    
    \textbf{1.2} & \textbf{Definition of central constructs related to maturity and maturation} \\
    & a) Maturity and dimensions of maturity & \cYes & \cYes & \cYes & \cYes \\
    & b) Maturity levels and maturation paths & \cYes & \cPart & \cNo & \cYes \\
    & c) Available levels of granularity of maturation & \cYes & \cPart & \cNo & \cYes \\
    & d) Underpinning theoretical foundations with respect to evolution and change & \cNo & \cYes & \cYes & \cNo \\
    
    \textbf{1.3} & \textbf{Definition of central constructs related to the application domain} & \cNo & \cYes & \cYes & \cNo \\
    
    \textbf{1.4} & \textbf{Target group-oriented documentation} & \cNo & \cPart & \cPart & \cNo \\
    \catrow{Descriptive}
    \textbf{2.1} & \textbf{Intersubjectively verifiable criteria for each maturity level and level of granularity} & \cNo & \cNo & \cPart & \cNo \\
    
    \textbf{2.2} & \textbf{Target group-oriented assessment methodology} & \\

    & a) Procedure model & \cNo & \cNo & \cPart & \cNo \\
    & b) Advice on the assessment of criteria & \cNo & \cNo & \cPart & \cNo \\
    & c) Advice on the adaptation and configuration of criteria & \cPart & \cNo & \cPart & \cNo \\
    & d) Expert knowledge from previous application & \cNo & \cNo & \cPart & \cNo \\

    \catrow{Prescriptive}
    \textbf{3.1} & \textbf{Improvement measures for each maturity level and level of granularity} & \cNo & \cNo & \cNo & \cNo \\
    
    \textbf{3.2} & \textbf{Decision calculus for selecting improvement measures} & \cNo & \cNo & \cNo & \cNo \\
    
    \textbf{3.3} & \textbf{Target group-oriented decision methodology} & \cNo &  \cNo & \cNo & \cNo \\
    \bottomrule
\end{tabularx}

\end{table}
\end{landscape}
}

\subsection{Divergence by Intended Purpose (Basic Principles)}

We found that each model only fulfills those basic principles most important for its intended purposes and target groups. We group the basic principles into four clusters below. The differences reflect the models' origins: a \gls{kpi} framework (\abbratis{}), a \csf{} community profile (\abbrdeloitte{}), and a sector-specific extension of the \camm{} (\abbrfsisac{}).

\subsubsection*{Positioning and basic information (DP~1.1a--e).}
The basic information requested by these \DPs{} frames a model's purpose and applicability, enabling users to choose the right model for their context. They include the \enquote{application domain} (and per \textcite{roglinger2012maturity} a well-defined \enquote{scope}), \enquote{prerequisites} for applying the model, the \enquote{purpose of use}, the \enquote{target group}, the \enquote{class of entities under investigation}, and a \enquote{differentiation from related maturity models} (positioning).

Developed as a financial-sector extension of the \camm{}, \abbrfsisac{} addresses a concrete community and correspondingly gives the most clearly defined application domain and target group (\enquote{executives and business leaders} and the technologists who operate the systems), even if its claim of cross-industry applicability broadens that domain again. That said, a target group spanning executives to technologists excludes almost no one, so naming a target group fulfills DP~1.1c without guaranteeing that each group is actually served; DP~1.4 probes this separately. \abbrdeloitte{}, by contrast, inherits the deliberate domain neutrality of its parent, the \nist{} \csf{}~2.0: as a community profile, it places no restrictions on sector or organization size, so its domain remains intentionally broad. \abbratis{}, a standalone \gls{kpi} framework centered on measurement, specifies neither domain nor target group tightly, focusing on the metrics rather than the framing. All three leave the entities under investigation vague, citing examples but not providing a structured overview. Notably, none of the models state explicit prerequisites. The lone hint is \abbrdeloitte{}'s reliance on \nist{} \csf{} 2.0, which presumes familiarity with that framework.

All three evaluated models claim to help evaluate, measure, or understand an organization's crypto agility or cryptographic resilience, a clear descriptive purpose. Furthermore, they promise to help \enquote{become crypto agile}, \enquote{meet the crypto agility KPIs}, or \enquote{advance cryptographic resilience}, a prescriptive purpose. %

\glsreset{caraf}
\abbrfsisac{} and \abbrdeloitte{} clearly position themselves towards existing models. \abbrfsisac{} \enquote{expands on [the \camm{}~\cite{hohmCamm2023}] by incorporating the technical aspects described in the model (\dots) and extends it by including process and other organizational aspects}. Additionally, it explicitly positions itself relative to other industry reports on crypto agility implementation, such as \cite{schmiegAgility2024,ibmcryptoagility2024}. \abbrdeloitte{}, while less detailed in this dimension, explicitly positions itself as a \nist{} \csf{} community profile and cites the \enquote{CISA Zero Trust Maturity Model 2.0}. \abbratis{}, developed independently, only refers to the \caraf{}~\cite{caraf2021} and acknowledges it as one approach to developing a maturity model. However, it does not explain how its approach differs from other models.

\subsubsection*{Design process and empirical validation (DP~1.1f).}
Documentation of the design and validation processes lets model users assess a model's robustness. The extent to which a model has been applied in practice is especially important for judging whether it fits an organization. As \textcite{wolfEval2025} noted, the \camm{} documents its design process and validation well. We attribute this to its origin in an academic setting, where peer review rewards explicit documentation. The three industry models, practitioner deliverables rather than peer-reviewed artifacts, do not document either. \abbrfsisac{} and \abbrdeloitte{} name the models they build on, but none describe how they were designed or validated, and only \abbrfsisac{} lists the contributors involved, which reveals little about the design process.

\subsubsection*{Maturity constructs and maturation logic (DP~1.2a--d).}
The goal of these design principles is to let users determine what constitutes maturity in the context of the model. They require structural properties, namely, defining maturity and its dimensions, and a set of maturity levels with a progression path. Moreover, they require the ability to read maturity at multiple levels of granularity (e.g., from high-level indicators for leadership down to fine-grained, per-dimension detail for technicians) and a theoretical underpinning of how maturity typically develops, explaining why maturation unfolds as the model claims.

The \camm{}, as the only academic model, does well in defining its levels and communicating granularity~\cite{wolfEval2025}. However, we find that for the industry models, the foundation a model builds on (if present) and its intended purpose largely determine the extent to which the structural properties are fulfilled. While all models define maturity as the achievement of crypto agility or cryptographic resilience, \abbrfsisac{} and \abbrdeloitte{} inherit their respective maturity dimensions from their foundational frameworks, whereas \abbratis{}, lacking such a foundation, defines its own dimensions.

Since \abbrfsisac{} directly builds upon the \camm{}, it inherits its maturity levels. However, its extended framework is not designed with the same level of detail, leaving the relation between framework steps and maturity levels open. \abbrdeloitte{} inherits structure and progression logic from its base, the \nist{} \csf{}, only extending it with a new \enquote{Tier 0: Basic}. \abbrdeloitte{} provides the clearest support for multiple levels of granularity: it comprises five top-level tiers that represent the achieved maturity level and can serve as a single data point to quickly and concisely communicate the status of maturation to leadership and executives. It also includes functions, each with categories subdivided into sub-categories, allowing users to drill deeper into each dimension. \abbratis{}, meanwhile, lacks the robustness of a base model to fall back on. While its \gls{kpi} scale could be interpreted as a set of maturity indicators, the corresponding metrics often lack explicit thresholds for determining maturity, and the indicators are very fine, lacking a higher-level score derived from their constituents or any global maturation levels.

The theoretical foundation, on the other hand, is unsurprisingly better presented by the comprehensive, practitioner-focused models \abbrfsisac{} and \abbratis{}. \abbrfsisac{} not only cites historical patterns in cryptographic transitions, the quantum computing threat, and regulatory pressure, but it also presents \enquote{Nine Core Elements of a Successful Crypto Agility Transition} outlining the major drivers of maturation. \abbratis{} likewise provides historical context and motivation, including a comparatively thorough introduction to (asymmetric) cryptography and quantum-related threats. Furthermore, the underlying drivers of change in an organization regarding crypto agility are discussed. \abbrdeloitte{}, as a terse conformance profile, describes no such aspects.

We find that these two properties vary independently. Structure can be inherited from a base model, though extending one, as \abbrfsisac{} does with the \camm{}, can be lossy. A theoretical foundation comes primarily from context-rich embedding in a document. FS-ISAC combines extensive background material with \camm{}'s structure, although the extension only partially preserves the base model’s maturity levels and granularity. \abbrdeloitte{} inherits the \csf{}'s structure but offers little crypto-agility-specific explanation. \abbratis{} provides such explanations but lacks explicit maturation levels and aggregated scores.

\begin{figure}[!t]
    \centering
    \input{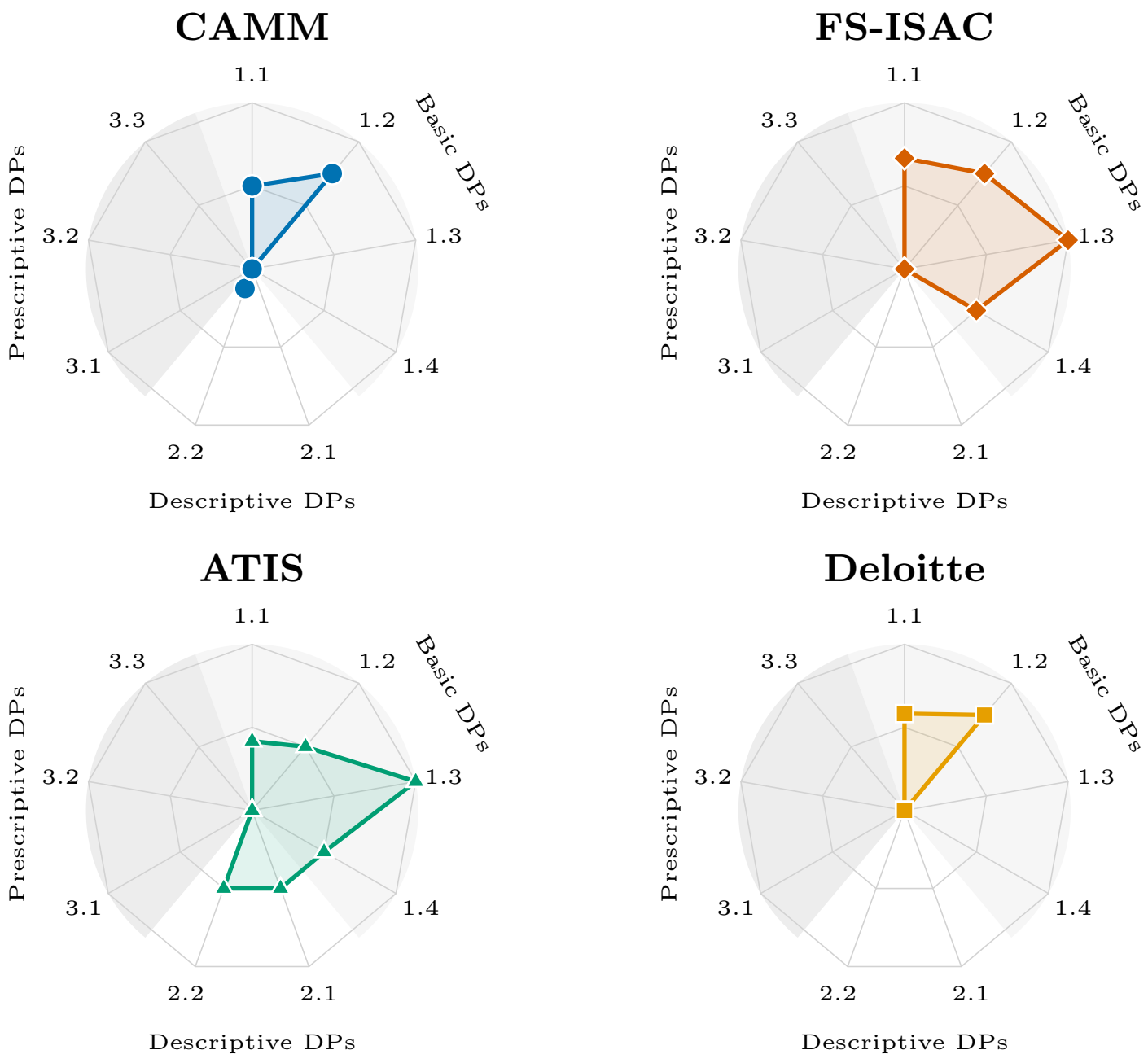}
    \caption{Per-model radar showing the mean fulfillment of the design principles averaged over their sub-items (\xmark{}$=0$, \mmark{}$=0.5$, \cmark{}$=1$). Sub-principles are equally weighted for simplicity. \textcite{poppelbuss2011} did not intend the principles as a combined score, so \cref{tab:results} remains the authoritative result. Results for the \camm{} are sourced from~\cite{wolfEval2025}.}
    \label{fig:model-profiles}
\end{figure}
\subsubsection*{Domain constructs and target group-oriented documentation (DP~1.3, 1.4).}
These principles concern whether the target group can understand the domain and how to apply the model within it. DP~1.3 requires clear, consistent definitions of the central elements of the application domain (crypto agility). DP~1.4 requires documentation that is accessible, comprehensible, and sufficiently detailed for the intended target groups to use the model. Following \textcite{roglinger2012maturity} and \textcite{wolfEval2025}, we assess accessibility and comprehensibility, as well as whether each target group is addressed sufficiently.

\abbrfsisac{} and \abbratis{} provide more crypto-agility-specific background than \abbrdeloitte{}. They provide an executive summary or overview for a leadership audience, along with substantial contextual detail for practitioners, and they describe the application domain in detail. At first glance, \abbrfsisac{} does a good job of target group-oriented documentation, with separate sections addressing different target groups and referencing other documents that provide guidance on specific aspects of crypto agility from the same authoring group. However, neither achieves full target group coverage, especially on what it takes to reach each maturity level. \abbrfsisac{} describes the (largely inherited) level requirements even more briefly than \camm{}, which \textcite{wolfEval2025} already found insufficient, while \abbratis{} lacks the details that technical staff need to assess some of its metrics.

\abbrdeloitte{}'s deliberate brevity leaves little room for extensive descriptions of domain constructs and documentation in general, though it inherits some information from the \csf{}. That \camm{} fails these design principles may be due to the inherent space constraints of academic publications, though the additional space permitted on the accompanying website~\cite{cammweb} weakens this interpretation.

These differences largely follow from the artifacts' purposes and foundations, as the distinctive profiles in \Cref{fig:model-profiles} show. The descriptive and prescriptive shortcomings discussed next do not.

\subsection{Verifiability is Systematically Underdeveloped (Descriptive Principles)}
The descriptive design principles are intended for maturity models that aim to describe and measure a system's current maturity. According to \textcite{poppelbuss2011}, this requires intersubjectively verifiable measurement criteria and target group-oriented assessment methodology. We found that for both \abbrfsisac{} and \abbrdeloitte{}, the assessment criteria are not intersubjectively verifiable, and no target group-oriented assessment methodology is provided, thereby limiting their descriptive value. \textcite{wolfEval2025} detected the same for the \camm{}, stating that several aspects need interpretation from assessors and thus risk different outcomes from different assessors, a weakness that \abbrfsisac{} inherits. \abbrdeloitte{}, as a CSF community profile, primarily relies on qualitative criteria. For example, GV.RR-03 requires that \enquote{adequate resources} be allocated, but neither adequacy nor the evidence required to demonstrate it is defined. Consequently, different assessors may reach different conclusions for the same organization.

The comparative standout in this category is \abbratis{}. Its 1--10 measurement scales provide discrete ratings and example scorecards with comments that offer limited guidance for interpreting and assessing the \glspl{kpi} as target group-oriented advice. Directly measurable criteria limit the room for different interpretations, and the provided examples can be useful for calibrating the endpoints of the scales. Even so, not all of these scales are intersubjectively verifiable: only 5 of 23 metrics provide an explicit transformation to the intended 1--10 scale. The rest rely on undefined judgment or give raw metrics with no mapping to the scale (e.g., \enquote{Assign a numerical score (e.g., 1 to 10) based on the assessed risk associated with implementing a new algorithm.}). Moreover, \abbratis{} does not clearly distinguish between the terms \kpi{} and metric, potentially leading to confusion among the target group. 

Thus, \abbratis{} is the model that fares best on the descriptive principles. That even the one model most oriented towards measurement still cannot consistently supply intersubjectively verifiable criteria suggests the gap is systematic.

\subsection{Improvement Guidance is Not Integrated (Prescriptive Principles)}

All models under test claim that they aim at providing guidance for improving crypto agility or cryptographic resilience. For maturity models intended for such prescriptive use, \textcite{poppelbuss2011} specify prescriptive design principles requiring improvement measures across maturity levels and levels of granularity, a decision calculus for selecting appropriate measures, and a target group-oriented decision methodology. As we found that the evaluated models universally fail these principles, we consolidate them in our analysis. Most importantly, none of the models directly integrate recommendations into the model itself, but the framing documents, in some cases, provide limited guidance.

\abbrfsisac{} provides a high-level eight-phase framework for cryptographic transitions in general, as well as \enquote{Nine Core Elements of a Successful Crypto Agility Transition}, which could be construed as improvement measures. The \abbrfsisac{} working group also refers to a separate survey of risk models~\cite{fsisacRiskModel}, advising readers to prioritize the riskiest assets. Beyond this high-level, risk-based prioritization, however, it provides no calculus for selecting among improvement measures and, apart from a single note linking one framework phase to a maturity level, keeps the measures separate from the model.

\abbratis{} is restricted to an attached one-page list of \enquote{recommendations}. It could be argued that improving the \glspl{kpi} would lead to better crypto agility, but this is not made explicit in the document, and prioritization lies solely with the users.

Among the tested models, \abbrdeloitte{} comes closest in prescriptive guidance: it offers qualitative prioritization guidance for cybersecurity risks (GV.RM-06) and for assets to migrate first (ID.AM-05), both inherited from the \csf{}. More broadly, the \nist{} \csf{}, with its tiers, in combination with an organization's profile, can be used to derive improvement measures, as shown, for example, in the prioritization process of \textcite{tang2026transit}. However, \abbrdeloitte{} offers no decision calculus (no weighting/scoring rule) and no link to the maturity tiers, thereby failing these principles.

\section{Discussion}

We observe that none of the three models explicitly positions itself as a formal maturity model. Their stated purposes range from presenting a \enquote{CSF 2.0 Community Profile} to outlining an approach for \enquote{Building Cryptographic Agility} in a specific sector. 
Nevertheless, NIST cites these documents in the context of crypto agility maturity models, suggesting that they are interpreted as such within the broader community.
We therefore evaluate them accordingly. At the same time, as \textcite{becker2009developing} stated, \enquote{A maturity model consists of a
sequence of maturity levels for a class
of objects. It represents an anticipated,
desired, or typical evolution path of these
objects shaped as discrete stages.}
Our evaluation reveals that \abbratis{}, in particular, lacks a precise definition of maturity levels. 
Therefore, according to \textcite{becker2009developing} in its current state \abbratis{}
 would require significant additions or modifications to meet the strict definition of a maturity model~\cite{becker2009developing}.

Additionally, one may ask whether formal maturity, in the sense of an evolutionary path shaped by discrete stages~\cite{becker2009developing}, is the right paradigm for crypto agility. \textcite{rameshan2026assessment} argue against such a single linear hierarchy for the application level and instead decompose agility into orthogonal dimensions. Our findings are consistent with this view, since part of the difficulty that all evaluated models have in defining intersubjectively verifiable criteria may stem from trying to force a multi-dimensional capability into a linear progression. The same multi-dimensionality plausibly holds at the organizational level: an organization strong on cryptographic inventory but weak on algorithm agility is not necessarily more or less agile than the reverse. Future model developers should therefore explicitly decide whether organizational crypto agility is best represented by a sequence of maturity levels or by several independent dimensions.

Moreover, the analyzed maturity models largely overlook the potential trade-offs associated with implementing extensive crypto agility.
While \abbrfsisac{} dedicates a section to the challenges of achieving crypto agility, none of the models address the risks an organization exposes itself to when undertaking such an endeavor. Risks highlighted in prior work~\cite{schmiegAgility2024}, such as heightened system complexity, larger attack surface, and increased maintenance, are not explicitly incorporated into the maturity assessments.
Decisions regarding crypto agility implementation must balance anticipated security benefits against their costs. Integrating such trade-off considerations into maturity models could enhance their practical relevance and their capacity to support decision-making.

\subsection{Considerations for Building Crypto Agility Maturity Models}
\label{sec:desiderata}

As \cref{sec:findings} shows, the strengths and shortcomings we identified are structural: fulfillment of the basic design principles follows largely from each model's intended purpose, whereas all fail the descriptive and prescriptive principles.

Since the fulfillment of basic principles reflects purpose rather than quality, a model should clearly state its purpose, including its application domain, scope, and target group. \abbrfsisac{} exemplifies this by giving the clearest application domain and target group of the models we studied, whereas even the academic \camm{} leaves its purpose implicit~\cite{wolfEval2025}. A clear purpose statement also helps prospective users choose the right model (see \cref{sec:selection}). The following four recommendations address crypto-agility-specific problems identified in our comparison; they complement the general design principles proposed by \textcite{poppelbuss2011} and \textcite{roglinger2012maturity}.

\subsubsection{Build on an established base, but carry its rigor through.}

A completely new design may sometimes be appropriate. However, our comparison suggests that models built on mature foundations generally exhibit stronger structural properties, provided the base's rigor is carried over. A well-integrated foundation provides structure, theoretical background, and often well-established documentation. Moreover, if the target group is already familiar with the foundation, they will find it much easier to apply the model. For example, \abbrdeloitte{} benefits from the structure, granularity, and qualitative prioritization guidance of the \nist{} \csf{}. This brings it closest to integrating improvement guidance (even though it, too, ultimately fails the prescriptive \DPs{}).

However, \abbrfsisac{}'s example shows that an extension still requires care. It builds on the \camm{}, but its design does not maintain the \camm{}'s rigor, and the model consequently performs worse on maturity levels and granularity. Our advice here is to (1) use a foundational model where it fits the new model's purpose, (2) clearly mark any extensions to the base, and (3) make sure extensions follow the same rigorous methodology as the base.

\subsubsection{Provide verifiable criteria.}

\Cref{sec:findings} shows that verifiable criteria are systematically underdeveloped. Even \abbratis{}, the model most oriented towards measurement, provides explicit formulas for 5 of its 23 metrics. Model developers can learn most from how those five differ from the rest: each is a count over a defined denominator (e.g., \enquote{Number of successful algorithm implementations / total number of attempted algorithm implementations}~\cite[Metric~12]{atisAgility2024}). However, even these formulas do not guarantee intersubjective verifiability when terms such as \enquote{successful}, \enquote{informed}, or \enquote{applicable} remain open to interpretation.

Of the remaining metrics, eight require an expert or assessor rating without a scoring rubric, while ten produce raw numbers not mapped to the 1--10 scale. For example, implementation time is objectively measurable, but its mapping is limited to the directional anchors \enquote{1 is a very long time and 10 is short}~\cite[Metric~1]{atisAgility2024}. Model developers should therefore define both the measurement inputs and the transformation from those inputs to assessment results. Where objective thresholds cannot reasonably be provided, models should state this limitation rather than presenting subjective judgments as reproducible scores. Other candidates for operationalized metrics include cryptographic-inventory coverage and migration progress per system.

These gaps illustrate a recurring measurement challenge in assessing crypto agility. Many relevant properties, such as an organization's understanding of trade-offs during cryptographic transitions, are inherently difficult to measure. Future models should distinguish explicitly between objective measurements, rule-guided expert judgments, and qualitative considerations rather than representing all three through an apparently uniform numerical scale.

\subsubsection{Integrate migration guidance.}

As \cref{sec:findings} concludes, all models present themselves as instruments for improving crypto agility, yet none integrates such guidance into the model itself. \textcite{poppelbuss2011} see the prescriptive value of a maturity model in the integration of a progression path into the maturity levels, with a calculus for choosing among measures.

For crypto agility in particular, a model could prioritize which cryptographic assets and algorithms to transition first on an explicit basis, such as risk exposure, asset criticality, and feasibility, and tie that ordering to its maturity levels, so that advancing a level corresponds to concrete migration milestones and an assessment result names the next step. \abbrfsisac{} advises addressing the riskiest assets first~\cite{fsisacRiskModel} and \abbrdeloitte{} inherits asset and risk prioritization (ID.AM-05, GV.RM-06). However, no evaluated model translates such advice into an integrated maturation path.

\subsubsection{Serve both leadership and technical audiences.}

Implementing crypto agility involves both decision-makers who allocate resources and technical staff who inventory, modify, and test cryptographic systems.
According to \textcite{poppelbuss2011}, a model should account for this by providing audience-matching granularity (DP~1.2c) and target group-oriented documentation (DP~1.4). As \cref{sec:findings} shows, these are independent requirements: \abbrdeloitte{} provides a five-tier headline that condenses to a single maturity rating, with functions and subcategories to drill into, but thin domain content to fill this structure. \abbratis{} has the opposite issue, since it offers fine-grained per-metric detail that never aggregates into a headline for leadership.

For model developers, the lesson is to combine these strengths into a single model. Include a concise headline indicator that leadership can use to authorize the budget and accept the residual risk of a migration. But also include per-dimension details concrete enough for technologists, and target group-oriented documentation with detailed examples at each level. This comes at the price of larger documents and more development effort, which we consider justified.

\subsection{Selecting a Crypto Agility Maturity Model}
\label{sec:selection}
Model choice depends on the intended assessment purpose and the users’ background. The design principles should therefore not be treated as an unweighted checklist: a model that fulfills more principles is not necessarily more suitable for a particular organization. In what follows, we offer some advice on model choice. Note, however, that a good fit can only be achieved by matching the basic principles to the intended purpose, since no model excels at descriptive measurement or prescriptive guidance.

\subsubsection{Choosing by purpose of use and background.}

The evaluated models differ primarily in which basic principles they fulfill, depending on the purpose they serve and the audiences they address. Only \abbrfsisac{} is sector-specific (though not restricted to it), making it a natural fit for the finance sector. For the other models, we suggest choosing based on the primary users' backgrounds.
If they already employ the \nist{} \csf{}, \abbrdeloitte{} will be easier to integrate into their governance structure than other models would be. The compromise is that the model provides little crypto-agility-specific depth and documentation.
For users comfortable with \glspl{kpi}, \abbratis{} offers the most fine-grained model but provides no clear maturity levels to communicate to leadership.
\abbrfsisac{} can serve both non-experts who need theoretical grounding and those already familiar with crypto agility, for different reasons: its framing document offers extensive documentation, and the academic foundation (only slightly loosened by its extension) provides solid coverage of relevant aspects of crypto agility.

\subsubsection{No good options for descriptive and prescriptive use.}
No evaluated maturity model provides clear guidance on improving crypto agility. We believe descriptiveness is a prerequisite for such guidance: making crypto agility measurable makes progress trackable. The goal is not to quantify everything but to ensure that different assessors reach the same verdict. The model under test that comes closest to being descriptive is \abbratis{}, since we consider 5 of the 23 metrics to be intersubjectively verifiable. For the more subjective measures, organizations can compensate by implementing their own scoring rubrics. Such adjustments will allow intra-organizational reproducibility, but not necessarily comparison between organizations. Similarly, for now, organizations will have to develop their own prioritization of which aspects of crypto agility to improve.

\subsection{Limitations and Future Work}

Our evaluation results were achieved through structured discussion between two researchers, with occasional tie-breaking input from a senior researcher. Although agreement was reached directly in most cases, the conclusions are ultimately subjective. To increase transparency, we document our verdicts in \cref{sec:findings} and report complete ratings in \cref{tab:results}. For detailed, per-principle explanations, we refer to our supplemental material~\cite{per_principle_eval}.

Furthermore, the authors of this work are researchers and do not necessarily belong to the models' primary target groups. While we consider the Design Principles by \textcite{poppelbuss2011} suitable for a structured assessment, certain aspects---e.g., the target group-oriented assessment methodology (\DP{}~2.2)---can ultimately only be validated through application in real-world organizational settings by the target group. Our evaluation, therefore, relies on reasoned assumptions regarding the needs of the intended users. As a result, our findings may over- or underestimate the practical applicability of the models and should be interpreted accordingly.

Finally, we decided not to complement our design-principle-based analysis with empirical application of the models. Given our finding that the models' assessment criteria are largely not intersubjectively verifiable (\DP{}~2.1), applying the models will yield primarily subjective results, reproducing our findings rather than independently testing them. The practical evaluation by \textcite{wolfEval2025} illustrates this: their case study resurfaced the same ambiguities as the design-principle-based analysis. As an example, applying \abbratis{} would require scoring metrics such as a hardware's resistance to side-channel attacks, for which the model prescribes only an \enquote{expert evaluation \dots{} (scale 1-10)}~\cite[Metric~23]{atisAgility2024} without any scoring rubric. Therefore, the empirical application of maturity models is useful, particularly for the target group, to validate aspects such as \DP{}~2.2. However, it requires a minimum of descriptiveness to yield relevant results.

An obvious next step for future work would be to incorporate our findings into a new crypto agility maturity model. To do so, more research is needed especially with respect to measurability of crypto agility criteria. 
While our focus was to evaluate the maturity models, the process of the evaluation with the Design Principles also led to discussions about the Design Principles itself. If they could be improved and refined for crypto agility maturity models is also left to future work.

\section{Conclusion}

In this paper, we evaluated three industry-driven crypto agility maturity models referenced by NIST against established design principles for maturity models~\cite{poppelbuss2011}. We find that the models meet basic design principles to varying degrees, which we mainly attribute to their intended purposes. However, they systematically fail to provide intersubjectively verifiable assessment criteria (descriptive principles) and integrated guidance for improving crypto agility (prescriptive principles) (RQ1, RQ2). Based on these findings, we offer practitioners advice on choosing among the models and derive concrete recommendations for developers of future crypto agility maturity models (RQ3).

If crypto agility maturity models are to inform governance mechanisms such as procurement requirements, regulatory guidance, or large-scale PQC migration strategies, future iterations must integrate measurable, reproducible criteria and explicit decision logic that links maturity assessment to concrete improvement paths.
Without such explicit criteria and decision rules, the models can organize discussion but cannot support reproducible assessments or determine the next migration step.

\subsubsection{Acknowledgements}
This work was supported by the National Research Center for Applied Cybersecurity ATHENE and the Dieter Schwarz Foundation.

\printbibliography

@techreport{tang2026transit,
  title={Transit Cybersecurity Framework Community Profile (Initial Public Draft)},
  author={Tang, CheeYee and Alshtein, Alex and Hardison, Matt and Holt, Emma and Sames, Christina and Tran, Hillary and others},
  year={2026},
  institution={National Institute of Standards and Technology},
  doi={10.6028/NIST.IR.8576.ipd}
}

@misc{nistConsiderations2025,
  title = {Considerations for {{Achieving Crypto Agility}}: {{Strategies}} and {{Practices}}:},
  shorttitle = {Considerations for {{Achieving Crypto Agility}}},
  author = {Barker, Elaine and Chen, Lidong and Regenscheid, Andrew and Moody, Dustin and Newhouse, William and Kent, Karen and Barker, William and Cooper, David and Souppaya, Murugiah and Housley, Russell and Turner, Sean},
  year = {2025},
  number = {39},
  institution = {{National Institute of Standards and Technology}},
  doi = {10.6028/NIST.CSWP.39},
  langid = {english}
}

@misc{fsisacAgility2024,
  title = {Building {{Cryptographic Agility}} in the {{Financial Sector Effective}}, {{Efficient Change}} in a {{Post Quantum World}}},
  author = {{Financial Services Information Sharing and Analysis Center ({FS-ISAC})}},
  shortauthor = {{FS-ISAC}},
  url = {https://www.fsisac.com/hubfs/Knowledge/PQC/BuildingCryptographicAgilityInTheFinancialSector.pdf},
  year = {2024}
}

@misc{deloitteAgility2025,
  title = {Cryptographic {{Resilience Community Profile}} ({{April}} 2025 {{Initial Draft}})},
  author = {Deloitte},
  year = {2025},
  url = {https://www.deloitte.com/content/dam/assets-shared/docs/services/consulting/2025/deloitte-cryptographic-resilience-community-profile-april-2025.pdf},
}

@inproceedings{hohmCamm2023,
author="Hohm, Julian
and Heinemann, Andreas
and Wiesmaier, Alexander",
title="Towards a Maturity Model for Crypto-Agility Assessment",
booktitle="Foundations and Practice of Security",
year="2023",
publisher="Springer Nature Switzerland",
isbn="978-3-031-30122-3",
doi="10.1007/978-3-031-30122-3_7"
}

@inproceedings{poppelbuss2011,
  title = {{What Makes a Useful Maturity Model? A Framework of General Design Principles for Maturity Models and its Demonstration in Business Process Managment.}},
  author = {Pöppelbu{\ss}, Jens and Röglinger, Maximilian},
  year = {2011},
  langid = {english},
  booktitle = {ECIS Proceedings. 28},
  url={http://aisel.aisnet.org/ecis2011/28}
}

@misc{atisAgility2024,
  title = {Strategic {{Framework}} for {{Crypto Agility}} and {{Quantum Risk Assessment}}},
  author = {ATIS},
  url = {https://atis.org/resources/strategic-framework-for-crypto-agility-and-quantum-risk-assessment/},
  year = {2024},
  langid = {american}
}

@inproceedings{wolfEval2025,
  title = {Practical {{Evaluation}} of the {{Crypto-Agility Maturity Model}}},
  author = {Wolf, Leonie and Umezulike, Samson and Öndarö, Gurur and Schinzel, Sebastian and Ising, Fabian},
  booktitle = {18th International Symposium on Foundations \& Practice of Security (FPS)},
  year = {2025},
  langid = {english},
  doi = {10.1007/978-3-032-20018-1_24}
}

@article{caraf2021,
	title = {{CARAF}: Crypto Agility Risk Assessment Framework},
	volume = {7},
	issn = {2057-2085},
	doi = {10.1093/cybsec/tyab013},
	shorttitle = {{CARAF}},
	number = {1},
	journal = {Journal of Cybersecurity},
	shortjournal = {J Cyber Secur},
	author = {Ma, Chujiao and Colon, Luis and Dera, Joe and Rashidi, Bahman and Garg, Vaibhav},
	date = {2021-01-01},
    year=2021,
}

@inproceedings{nethenPQC2024,
	title = {{PMMP}-{PQC} Migration Management Process},
	isbn = {979-8-4007-1651-5},
	doi = {10.1145/3655693.3655719},
	eventtitle = {{EICC} 2024: European Interdisciplinary Cybersecurity Conference},
	booktitle = {European Interdisciplinary Cybersecurity Conference},
	publisher = {{ACM}},
	author = {Von Nethen, Nils and Wiesmaier, Alexander and Alnahawi, Nouri and Henrich, Johanna},
	date = {2024-06-05},
    year = 2024,
	langid = {english},
}

@article{looyEvaluating2017,
	title = {Evaluating Business Process Maturity Models},
	volume = {18},
	issn = {15369323},
	doi = {10.17705/1jais.00460},
	number = {6},
	journal = {Journal of the Association for Information Systems},
	shortjournal = {{JAIS}},
	author = {{Van Looy}, Amy and Poels, Geert and Snoeck, Monique},
	date = {2017-06},
    year=2017,
	langid = {english},
}

@article{roglinger2012maturity,
  title={Maturity models in business process management},
  author={R{\"o}glinger, Maximilian and P{\"o}ppelbu{\ss}, Jens and Becker, J{\"o}rg},
  journal={Business process management journal},
  volume={18},
  number={2},
  year={2012},
  publisher={Emerald Group Publishing Limited},
  doi={10.1108/14637151211225225}
}

@misc{nistcsf2024,
  title = {{The NIST Cybersecurity Framework (CSF) 2.0}},
  author = {{NIST}},
  %shortauthor = {{NIST}},
  year = {2024},
  doi = {10.6028/NIST.CSWP.29},
}

@misc{fsisacRiskModel,
    title = {{Risk Model Technical Paper}},
    author = {{PQC Working Group, FS-ISAC}},
    year = {2023},
    url = {https://www.fsisac.com/hubfs/Knowledge/PQC/RiskModel.pdf}
}

@misc{schmiegAgility2024,
    title = {{Cryptographic Agility and Key Rotation}},
    url = {https://bughunters.google.com/blog/cryptographic-agility-and-key-rotation},
    year = {2024},
    author={Sophie Schmieg and  Stefan Kölbl and Guillaume Endignoux}
}

@misc{dod-cmmc,
    url={https://www.acquisition.gov/dfars/252.204%2D7021%2Dcontractor%2Dcompliance%2Dcybersecurity%2Dmaturity%2Dmodel%2Dcertification-level-requirements%2E},
    date={2025-11-10},
    year=2025,
    title={252.204-7021 Contractor Compliance With the Cybersecurity Maturity Model Certification Level Requirements.},
    author={{U.S. Department of Defense (DoD)}}
}

@misc{chen_report_2016,
	title = {Report on {Post}-{Quantum} {Cryptography}},
	%howpublished = {\url{https://nvlpubs.nist.gov/nistpubs/ir/2016/NIST.IR.8105.pdf}},
	abstract = {In recent years, there has been a substantial amount of research on quantum computers –machines that exploit quantum mechanical phenomena to solve mathematical problems that are difficult or intractable for conventional computers. If large-scale quantum computers are ever built, they will be able to break many of the public-key cryptosystems currently in use. This would seriously compromise the confidentiality and integrity of digital communications on the Internet and elsewhere. The goal of post-quantum cryptography (also called quantum-resistant cryptography) is to develop cryptographic systems that are secure against both quantum and classical computers, and can interoperate with existing communications protocols and networks. This Internal Report shares the National Institute of Standards and Technology (NIST)’s current understanding about the status of quantum computing and post-quantum cryptography, and outlines NIST’s initial plan to move forward in this space. The report also recognizes the challenge of moving to new cryptographic infrastructures and therefore emphasizes the need for agencies to focus on crypto agility.},
	number = {NIST IR 8105},
	institution = {National Institute of Standards and Technology},
	author = {Chen, Lily and Jordan, Stephen and Liu, Yi-Kai and Moody, Dustin and Peralta, Rene and Perlner, Ray and Smith-Tone, Daniel},
	month = apr,
	year = {2016},
	doi = {10.6028/NIST.IR.8105},
}

@misc{eu-pqc-roadmap,
    title = {A Coordinated Implementation
Roadmap for the Transition to
Post-Quantum Cryptography},
    author={{EU PQC Workstream}},
    year=2025,
    url={https://digital-strategy.ec.europa.eu/en/library/coordinated-implementation-roadmap-transition-post-quantum-cryptography}
}

@misc{ibmcryptoagility2024,
    title={Crypto-agility and quantum-safe readiness},
    author={Ray Harishankar and Michael Osborne and Jai S. Arun and John Buselli and Jennifer Janechek},
    url={https://www.ibm.com/quantum/blog/crypto-agility},
    date={2024-06-19},
    year={2024}
}

@misc{BSI-TR-02102-1,
    title = {{BSI TR-02102-1} - Cryptographic Mechanisms:
Recommendations and Key Lengths},
    author={{Bundesamt für Sicherheit in der Informationstechnik (BSI)}},
    year=2026,
    url={https://www.bsi.bund.de/SharedDocs/Downloads/EN/BSI/Publications/TechGuidelines/TG02102/BSI-TR-02102-1.pdf}
}

@article{nather2024toward,
  title={Toward a Common Understanding of Cryptographic Agility--A Systematic Review},
  author={N{\"a}ther, Christian and Herzinger, Daniel and Stegh{\"o}fer, Jan-Philipp and Gazdag, Stefan-Lukas and Hirsch, Eduard and Loebenberger, Daniel},
  journal={arXiv preprint},
  year={2024},
  doi={10.48550/arXiv.2411.08781}
}

@article{natherMigrating2024,
    title = {Migrating Software Systems Toward Post-Quantum Cryptography-A Systematic Literature Review},
    author = {N{\"a}ther, Christian and Herzinger, Daniel and Gazdag, Stefan-Lukas and Stegh{\"o}fer, Jan-Philipp and Daum, Simon and Loebenberger, Daniel},
    year = {2024},
    doi = {10.1109/ACCESS.2024.3450306},
    journal = {IEEE Access (Volume: 12)},
}

@article{dikhanbayeva2020assessment,
  title={Assessment of Industry 4.0 Maturity Models by Design Principles},
  author={Dikhanbayeva, Dinara and Shaikholla, Sabit and Suleiman, Zhanybek and Turkyilmaz, Ali},
  journal={Sustainability},
  volume={12},
  number={23},
  year={2020},
  publisher={MDPI},
  doi={10.3390/su12239927}
}

@article{felch2020make,
  title={How to make business process maturity models better--drawing on design science research},
  author={Felch, Vanessa and Asdecker, Bj{\"o}rn},
  journal={Pacific Asia Conference on Information Systems (PACIS)},
  year={2020},
  url={https://aisel.aisnet.org/pacis2020/50/}
}

@article{ferraz_bim_maturity_models_2020,
  title = {BIM Maturity Models Evaluated by Design Principles},
  author = {Ferraz, Cristine and Loures, Eduardo Rocha and Deschamps, Fernando},
  year = 2020,
  journal = {Transdisciplinary Engineering for Complex Socio-technical Systems -- Real-life Applications},
  doi = {10.3233/ATDE200111},
  langid = {english}
}

@inproceedings{frick_identification_design_elements_2012,
  title = {Identification of {{Design Elements}} for a {{Maturity Model}} for {{Interorganizational Integration}}: {{A Comparative Analysis}}},
  booktitle = {Proceedings of the 25th {{Bled eConference eDependability}}: {{Reliable}} and {{Trustworthy eStructures}}, {{eProcesses}}, {{eOperations}} and {{eServices}} for the {{Future}}},
  author = {Frick, Norbert},
  date = {2012-06-17},
  year=2012,
  eventtitle = {25th {{Bled eConference eDependability}}: {{Reliable}} and {{Trustworthy eStructures}}, {{eProcesses}}, {{eOperations}} and {{eServices}} for the {{Future}}},
  url={https://aisel.aisnet.org/bled2012/10/},
  langid = {english}
}

@misc{erol_quantum_readiness_cryptography_2025,
  title = {Quantum {{Readiness}} in {{Cryptography}}: {{A Maturity-Based Framework}} for {{Post-Quantum Transition}}},
  shorttitle = {Quantum {{Readiness}} in {{Cryptography}}},
  author = {Erol, Volkan},
  date = {2025-10-02},
  year=2025,
  eprinttype = {Computer Science and Mathematics},
  doi = {10.20944/preprints202509.2584.v1},
  langid = {english},
  note={Preprint}
}

@article{jena_quantum_security_deadline_2025,
  title = {The {{Quantum Security Deadline}}: {{Building Crypto-Agility Against Harvest Now}}, {{Decrypt Later}}’ {{Threats}}},
  shorttitle = {The {{Quantum Security Deadline}}},
  author = {Jena, Jyotirmay},
  date = {2025-09-26},
  year={2025},
  journal = {European Journal of Computer Science and Information Technology},
  shortjournal = {EJCSIT},
  volume = {13},
  number = {52},
  issn = {20540957, 20540965},
  doi = {10.37745/ejcsit.2013/vol13n523552},
  %url = {https://eajournals.org/ejcsit/vol13-issue52-2025/the-quantum-security-deadline-building-crypto-agility-against-harvest-now-decrypt-later-threats/},
  abstract = {The threat from Cryptographically Relevant Quantum Computers (CRQCs) has evolved from a distant hypothesis into an urgent security reality. This article asserts that the true deadline for Post-Quantum Cryptography (PQC) migration is now, driven by the “Harvest Now, Decrypt Later” (HNDL) threat model—where adversaries exfiltrate encrypted, long-lived data today, anticipating its future decryption by quantum means. To counter this emerging risk, organizations must adopt the Crypto-Agility Mandate, a proactive architectural strategy designed to safeguard systems before CRQCs reach operational maturity. The proposed roadmap focuses on four immediate imperatives: conducting a comprehensive Cryptographic Bill of Materials (CBOM) to map existing encryption dependencies; deploying Hybrid Cryptography to bridge classical and quantum-safe algorithms; automating Certificate Lifecycle Management (CLM) to manage escalating cryptographic complexity; and enforcing PQC compliance across the digital supply chain. By embedding crypto-agility today, enterprises can fortify their digital infrastructure and ensure long-term resilience in the approaching quantum era.}
}

@misc{bsiSecuring2025,
    title = {Securing Tomorrow, Today: Transitioning to Post-Quantum Cryptography},
    author={{Bundesamt für Sicherheit in der Informationstechnik (BSI)}},
    year=2024,
    url={https://www.bsi.bund.de/SharedDocs/Downloads/EN/BSI/Crypto/PQC-joint-statement.pdf}
}

@misc{tnoPQCMigration2025,
    title = {The PQC Migration Handbook - Guidelines for Migrating to Post-Quantum Cryptography (2nd Edition)},
    author={Alessandro Amadori and Thomas Attema and Maxime Bombar and João Diogo Duarte and Vincent Dunning, Simona Etinski and Daniël van Gent and Matthieu Lequesne and Ward van der Schoot and Marc Stevens and {AIVD Cryptologists and Advisors}},
    year=2024,
    url={https://publications.tno.nl/publication/34643386/fXcPVHsX/TNO-2024-pqc-en.pdf}
}

@article{becker2009developing,
  title={Developing maturity models for IT management: A procedure model and its application},
  author={Becker, J{\"o}rg and Knackstedt, Ralf and P{\"o}ppelbu{\ss}, Jens},
  journal={Business \& information systems engineering},
  volume={1},
  number={3},
  year={2009},
  publisher={Springer},
  doi={10.1007/s12599-009-0044-5}
}

@misc{cammweb,
    author={{UCS Research Group}},
    title={Crypto-Agility Maturity Model ({CAMM})},
    year={2022},
    url={https://camm.h-da.io/},
    urldate={2026-06-01}
}

@misc{qramm2025,
  title = {{QRAMM}: The Quantum Readiness Assurance Maturity Model},
  author = {{CyberSecurity NonProfit (CSNP)}},
  date = {2025},
  url = {https://qramm.org/},
  urldate = {2026-07-17},
}

@misc{pqcmm2025,
  title = {Post-Quantum Cryptography Maturity Model ({PQCMM})},
  author = {{PKI Consortium PQC Working Group}},
  date = {2025-10-27},
  url = {https://pkic.org/wg/pqc/pqcmm/},
  urldate = {2026-07-17},
}

@inproceedings{rameshan2026assessment,
  title = {Cryptographic Agility for Applications: {{An}} Assessment Framework and Principled {{API}} Design},
  booktitle = {Migration and Agility in Cryptographic Systems},
  author = {Rameshan, Navaneeth and Messmer, Gregoire},
  date = {2026},
  publisher = {Springer Nature},
  doi = {10.1007/978-3-032-28946-9_9}
}

@misc{per_principle_eval,
    author = {{Anonymous Authors}},
    title={{Digital Appendix: Per-Principle Evidence for "Maturity$^2$}: Assessing the Maturity of Crypto Agility Maturity Models"},
    doi = {10.5281/zenodo.22892506},
}

\end{document}